\documentclass{article}

\usepackage{todonotes}
\usepackage{color, colortbl}
\usepackage{tcolorbox}
\definecolor{Gray}{gray}{0.9}
\usepackage{natbib}
\usepackage[T1]{fontenc}
\usepackage[utf8]{inputenc}
\usepackage{authblk}
\PassOptionsToPackage{hyphens}{url}\usepackage{hyperref}

\title{Digital Voodoo Dolls}

\author[1]{Marija Slavkovik\thanks{marija.slavkovik@uib.no}}
\author[2]{Clemens Stachl\thanks{stachl@stanford.edu}}
\author[3]{Caroline Pitman\thanks{pitmanc@cua.edu}}
\author[3]{Johnathan Askonas \thanks{askonas@cua.edu}}
\affil[1]{University of Bergen, Norway}
\affil[2]{Stanford University, USA}
\affil[3]{The Catholic University of America, USA}

\begin{document}

\maketitle

\begin{abstract}
An institution, be it a body of government, commercial enterprise, or a service, cannot interact directly with a person. Instead, a model is created to represent us. We argue the existence of a new high-fidelity type of person model which we call a digital voodoo doll. We conceptualize it and compare its features with existing models of persons. Digital voodoo dolls are distinguished by existing completely beyond the influence and control of the person they represent. We discuss the ethical issues that such a lack of accountability creates and argue how these concerns can be mitigated.  
\end{abstract}

\section{Introduction} % Jon take a crack
When asked to describe themselves, each person selects a somewhat different set of properties, attributes, and potentials that they think define them overall as a person. Most of these self-attributed features can change over time or could be described differently to people, dependent on the situation (e.g., job interview vs beers with friends). Past research suggests that people can judge some personal attributes better than others \citep{Vazire2010}. However, to some degree, people are generally able to shape their reputation based on their interests. 

What if your self-perception would not matter? What if who you are is not determined by yourself but an algorithm that determines your personality by using data on your past behaviors, choices, and desires? What if you cannot easily change what it "concludes" about you and who is getting access to this information?
We constantly create large amounts of data traces that provide a wide range of clues on who we really are. Information that not only reflects what we think we are and information that we want others to see about us. This data is used among else to build digital models of individuals, which enable service providers to personalize the service-user interaction experience. Person-specific models can be very simple and include just a role in a system: you get to see a different exam interface depending on whether you are logged in as a student or as a teacher. Due to the vast amounts of data available, latest computational capabilities technologies, and advances in artificial intelligence, it is now possible to generate complex, high-fidelity models of people. These new types of models not only contain descriptions of your features (e.g., shoe-size), but they also include indirect estimates of your propensities, traits, and potentially your vulnerabilities (e.g., emotional stability). 
Here, we call these models {\em digital voodoo dolls.} 

A digital voodoo doll is a dynamically generated information construct that models a person and their intentions. It is constructed from a collection of data created from digital traces, estimates derived by comparison with other data (from other people), and outputs from machine learning algorithms. 

 A digital voodoo doll can be created entirely without your permission or even awareness. For example, you can be assigned to the category \emph{woman} because you have clicked on the same or similar things typical women have clicked on. If those other women collectively start clicking on something else and you do not, you might just become a man. Any errors in the representation are also not within your immediate power to correct, nor will the model change if you decide to change your life.

Because digital voodoo dolls are dynamically generated digital constructs, they might not simply vanish when being deleted: as long as the ingredients exist, a new voodoo doll can be summoned. The ingredients for its creations are the data traces of a whole society, distributed in many copies, housed in online data warehouses, and on sale. As a consequence, the use of digital voodoo dolls transcends the typical ethical concerns regarding digital tracking and data collecting, personalization, and targeted advertising. Because the modeled person has no direct power over the voodoo doll's creation and use, the only viable way to change the digital voodoo doll behavior is to systematically change one's own behavior lastingly. This makes the digital voodoo doll also a possibly very powerful tool of behavioral control and manipulation. 
 
Here, we describe digital voodoo dolls and contrast them with other digital person-models of different fidelity. We argue that this new high fidelity model has features that pose new and unique dangers to the autonomy of people in a sociotechnical society. The goal of this article is to curb the impact of digital voodoo dolls before they become ubiquitous by elaborating what they are and what distinguishes them. We also discuss why the current approaches to curtailing threats from the datafication of society are not efficient on voodoo dolls.  

A well-developed concept is necessary to understand the ethical risks brought about by the unscrupulous use of digital voodoo dolls. We also hope that this analysis helps people recognize that what they are building is a voodoo doll and if their intention is to not contribute to the soft algorithmic oppression of people, they will modify their behavior.

%startbox
The idea that a digital or algorithmic identity of a person is indirectly created when aspects of our lives are turned into data has been discussed in the literature. Specifically, \citet{Cheney-Lippold2011} discusses algorithmic identities that we may be assigned beyond our control: "This force  is not entirely benign but is instead something that tells
us who we are, what we want, and who we should be. It is removed from traditional mechanisms for resistance and ultimately requires us to conceive
of freedom, in whatever form, much more differently than previously
thought. We are effectively losing control in defining who we are online, or
more specifically we are losing ownership over the meaning of the categories
that constitute our identities."
The earlier work of \citet{Elmer2003} focuses on consumer profiling and feedback technologies that shape our choices and indirectly, identity. ``Surveillant assemblages'' and ``data doubles'' as described by \citet{haggerty_surveillant_2000} and \citet{lyon_surveillance_2007} as the network of surveillance and information flows about an individual, amounting to a kind of software description or ``double'' of that user.

These terms are capacious: ``data doubles'' refers to information ranging from surveillance data on classified government systems to the personality one portrays : it might be understood as the sum total of the image of a person that might be gathered in reviewing all digitally available information anywhere. In practice, the term is often used to describe the models that arise in particular contexts, such as the ``data double'' of one's social media profiles that might be scanned by a prospective roommate. Similarly, ``algorithmic identity'' can be broadly understood as a philosophical, even metaphysical, shift in how individuals, organizations, and even states understand categories as such, shaping ``offline'' ideas in law, business, and self-conception.

Digital voodoo dolls have two distinguishing features when compared to Lyon's/Ruppert's ``data double'', Greg Elmer's ``profiles'', and Cheney-Lippold's  ``algorithmic identity''. The person model of the ``digital voodoo doll'' described in this paper is a special class of a data double, created by programatically networked data exchange and scoring, informing predictions of future behavior and desire. It is distinguished from these previous frameworks (or as a special sub-class of them) by two properties, which generate distinctive technical, legal, and ethical challenges: emergence and prediction.  They reflect not only our identity moulded under the pressure of datafication and surveillance, but exist as a time-dilated distorted mirror of who we are depending on the underlying data storage and architecture characteristics of the socio-technical systems. The digital voodoo doll model focuses not on digital identity or surveillance (in their myriad forms) but specifically on the kinds of person models produced by pragmatically interacting systems of data exchange. The second distinctive aspect of the digital voodoo doll model is the focus not only what the user appears to have done, but on the incorporation of this data into recommendation engines and predictive models (including social scoring systems). This latter aspect is only made recently possible with the advances of big data machine learning. Moreover, prediction requires only the discretization of features, and not necessarily the assignment of users into categories  \citep{Cheney-Lippold2011}. While alleviating issues of possible stereotyping or categorical misidentification, this raises additional ethical concerns we discuss.
%endbox

The main contribution of this work is the development of the concept of digital voodoo dolls. Creating and using these personal models has the potential to negatively affect people and softly push for a change of behavior in society. By clearly distinguishing digital voodoo dolls from other person models currently in operation we hope to initiate a discussion of how to avoid their negative side effects and empower people in today's surveillance-eased world. 

The paper is structured as follows. Digital voodoo dolls are made possible by the unprecedented amount of data about individuals that are generated daily. In Section  {\em Creation and collection of digital data} we give a brief overview of how this data is collected and what implications can be drawn about an individual from it. In Section {\em Modeling people}  we give a list of the existing models of people/individuals and their features. In Section {\em Digital Voodoo Dolls} we develop the concept of this new model and contrast it with the models discussed in the previous section. Section {\em Ethical concerns} we discuss the main values which can be impacted by the existence and use of digital voodoo dolls. Section {\em Possible paths to responsible voodoo } discusses two possible approaches by which a user can assert its influence on digital voodoo doll creation. In the last Section we summarise the work and outline directions for future work.

%In Section ``Digital footprint" %\ref{sec:footprint}
%we give an overview of which information about a person can be digitally accessed and how that information can be used to make inferences about that persons character and behaviour. In Section ``Modeling People'' %~\ref{sec:models}
%we discuss the existing methods used to build a user model in a sociotechnical system. In Section ``Digital Voodoo Dolls vs User Models'' %~\ref{sec:voodoo} 
%we contrast the introduced user models with our concept of digital voodoo doll with the purpose of positioning this new digital construct in the existing landscape of user-sociotechincal system interaction. In Section ``Ethical Concerns" %~\ref{sec:concern} 
%we discuss existing ethical concerns regarding user modeling in related work and analyse how those concerns extend to digital voodoo dolls.  In Section ``Dismantling voodoo dolls" we discuss possible approaches to mitigating the identified ethical concerns. 

\section{Creation and collection of digital data}\label{sec:footprint}%what voodoo looks like at a technical level
%\begin{itemize}
 %   \item what can be collected (sensors)
  %  \item what can be inferred  (patterns)
  %  \item what is user modeling \cite{Fischer2001}
  %  \item basics of cookies, trackers %\url{https://www.technologyreview.com/s/613996/youre-very-easy-to-track-down-even-when-your-data-has%-been-anonymized/}, device fingerprinting \url{https://iq.opengenus.org/audio-fingerprinting/}, %bio-metrics (short introduction and comparison) \cite{Acar:2014}
%\end{itemize}
%https://spreadprivacy.com/duckduckgo-tracker-radar/

We give a brief overview of the state of the art in the creation and collection of digital data, because it is this state of affairs that has made the existence of digital voodoo dolls possible. In this section, we (1) provide a short overview on how people passively generate data as part of their daily life, (2) how smartphones in particular can be used to actively collect fine-grained data about people's behavior and contexts, and (3) what kind of inferences can be made from these digital footprints.

The digitization of our daily lives has made it possible to do many things conveniently from any place with an internet connection. This development has however also facilitated the collection of data on individuals' actions, preferences, and indirectly their traits. Moreover, automated data collection has become the norm rather than the exception. The ubiquity of automated tracking systems has made it almost impossible to escape these practices \citep{Auxier2019}. 

The data we leave behind can be used to optimize products and services, to tailor advertisements to individuals, but also this data can be used to make predictions of our future behavior, preferences and our momentary states (e.g., emotional vulnerability, mood). Digital footprints have also been used to infer peoples personal traits and psychological dispositions \citep{Kosinski2013}. 

%Digital footprints
For a long time, what we do, when, where - digitally has been subject to tracking and online fingerprinting based on our most basic behaviors \citep{Ahmed2008}. A digital economy with a myriad of services and products has been built based on how people surf the web and which advertisements are best catered to them, at specific times, based on their past activity \citep{Acar:2014, ICO2019}. Websites we visit, queries we search for, how we type on our keyboard, what we like, and when we turn the page on our e-reader is tracked. These passively generated (e.g., opening a website) or actively left behind (e.g., posting a status update on Instagram) digital footprints are used to uniquely identify individuals (fingerprinting) and to personalize content based on their needs, interests, and desires. This process also leads to the creation of the most accurate digital representations of individuals and their characteristics (user models).

%Mobile Sensors
%Alongside the radical progress that has been made in computing-, sensor-, and connectivity technology, new and more fine grained methods for behavioral tracking have become available. 
One of the most radical changes in computing technology has been the shift to mobile, ubiquitous computing \citep{Abowd2000, IoT2004}, and the ongoing deployment of artificial intelligence to computing devices \citep{Taddeo2018}. Reflected by the mobile-first development approaches of many software companies, smartphones have become the new default computers \citep{Alghannam2018}. In that sense, harnessing the concentrated potential of on-board sensors and the computational power of smartphones to quantify individual characteristics is not as far-fetched as it might seem. Smartphone apps can easily be re-purposed to collect data on how people behave in different contexts \citep{Harari2018, Harari2016}. 
% Smartphones also carry a number of increasingly precise sensors that can measure the phone’s location (GPS, Wifi), orientation in 3D space (Gyroscope), and movement (accelerometer). Furthermore, smartphones can gather information on the immediate environment and the events that happen in close proximity to it. Auditory input can be measured via microphones, illumination can be retrieved from the ambient light sensor and the cameras. Connections with other technical devices and infrastructure can be informative about environmental characteristics. For example, Bluetooth and Wifi connections can inform about the phones current environment. Some phones are even able to measure humidity, atmospheric pressure, ambient temperature and radiation levels in their environment.

In addition to direct sensor output, the computational abilities of modern smartphones allow these systems to fuse input from different sensors to infer higher-level activities (e.g., "in a cafe" based on location and ambient noise). Naturally, smartphones can also be used to track the activities users perform on their phones. These activities include communication behavior \citep{Harari2019, Montag2015}, installation and use of apps \citep{Bohmer2011, Stachl2017}, music preferences \citep{Nave2018, Stachl2020}, keystrokes \citep{Buschek2015}, general day- and nighttime activity \citep{Schoedel2020}, and everything that is visible on the screen \citep{Ram2020}.

% Besides smartphones, many other, digital devices collect data that can inform about people's behavior and the environment they are in. Latest cars for example can measure how we drive and what music we listen to, wearables can measure our physical states, lawn mowers and vacuum cleaners can infer the size of our home, our toothbrush when and how often we clean our teeth. We could go on forever. While any of these data (digital footprints and behavioral data from devices) might not appear to sensitive by itself, an increasing number of studies shows that very personal inferences can be made by modeling these data with machine learning methods, in particular when combining them.

The easiest type of inferences that can be made from our digital footprints concerns the recognition of demographic characteristics (e.g., gender and age). Gender prediction for example, has been demonstrated on a wide range of different data-types, including websites visits \citep{Hu2007,Zhong2013}, social media \citep{Kosinski2013}, photos \citep{Ng2015}, gait \citep{Yu2009}, driving \citep{Stachl2015}, and language use \cite{Cheng2011}. While the automated recognition of gender might seem banal, it can have a significant real-world impact once sexual orientation is predicted from the data people leave behind \citep{Wang2018}\footnote{This paper has been challenged, on ethical and scientific grounds, by ethicists and by computer scientists (see for example \url{https://medium.com/@blaisea/do-algorithms-reveal-sexual-orientation-or-just-expose-our-stereotypes-d998fafdf477})}. More recent work suggests that facial features can even be used to estimate people's political orientation \citep{Kosinski2021}.

Moreover, the automatic prediction of peoples' personality trait levels from digital footprints and behavioral data has become a very active area of research. Personality traits are often defined as relatively stable patterns of thought, feelings, and behavior \citep{Matthews2009} and have been shown to predict a number of important life outcomes \citep{Ozer2006, Roberts2007, Soto2019}. Past research suggests that personality traits can be predicted from a range of data types including text \citep{Park2015}, Facebook Likes \citep{Youyou2015, Azucar2018}, spending records \citep{Gladstone2019}, and smartphone data \citep{Stachl2020} with some accuracy. 

Some past work has also aimed to predict individual differences in cognitive abilities, with less success \citep[i.e., intelligence, ][]{Gordon2019, Kosinski2013, Wei2017}. Finally, it has been shown that the data people leave behind in online social networks can be used to predict most sensitive psycho-pathological states and conditions \citep[i.e., depression and mood][]{Servia-Rodriguez2017, Eichstaedt2018}. For example, the language used on Facebook can be used to infer whether a person is depressive or not.

\section{Modeling people}\label{sec:models} %Jon to signpost
In this section we list different models used to model a person both offline and online. We do this to contrast these existing models with the new concept of digital voodoo doll we are constructing. 

\begin{table*}[h!]
	\centering 
\resizebox{\linewidth}{!}{%
\begin{tabular}{|c|c|c|c|c|c|}\hline
 \rowcolor{Gray}  &\bf Person-Institution Relationship & \bf Permanence & \bf Can create/delete & \bf Can edit & \bf Purpose\\\hline
\bf Identifying & one to one     & strong & institution & person + institution & assign rights and obligations\\ \hline
\bf User Account & one to many   & weak & person & person & interaction\\ \hline
\bf Verified Account &one to one & strong & person or institution & person & interaction \\\hline
\bf User Model & many-to-many & dynamic &  institution  & institution/person & personalization \\\hline
\bf Look-a-like Model & many-to-one & dynamic &  institution  & institution/person & targeted content \\\hline
\bf Digital Voodoo Doll &  one to one  & dynamic  & cross-institutional construct   & no one &  nudging/ option architecture  \\\hline
\end{tabular}
}\caption{Different models of people and their properties.} \label{tab:models}
\end{table*}

\begin{figure}
    \centering
\includegraphics[width=0.5\textwidth]{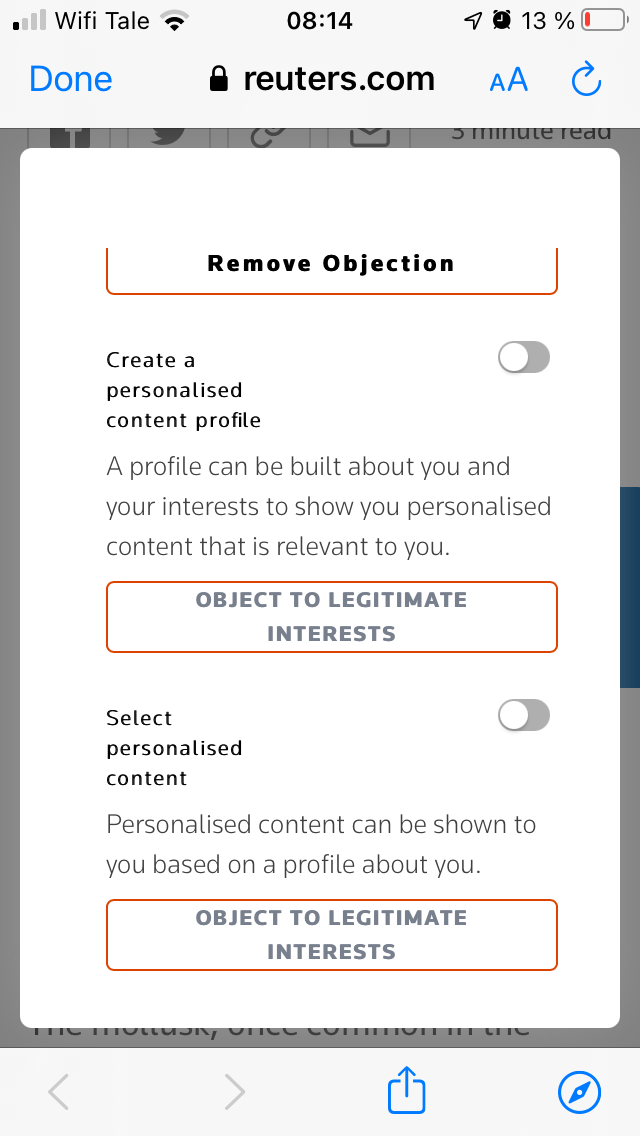}
    \caption{An example of how a person profile is constructed from data traces and used. Cookie consent pop-up from reuters.com grabbed May 6, 2021} \label{fig:example}
\end{figure}

People can interact directly with each other, however when a person needs to interact with an agent that is not a person, such as an institution, or a service, a model of the person is needed to execute that interaction. The modeling of people serves numerous institutional purposes including enabling the institution to make decisions regarding the people it interacts with. Figure~\ref{fig:example} gives an example of how a news outlet uses models of people who are engaging with it. 

In the most abstract sense, a model of a person is a data structure that is associated with that person created by an institution to serve a specified institutional purpose. For example, a record containing your name, your social security number, your date of birth and your current contact details is a model of you. Another example of a model of you is your social media profile. A social media platform has two profiles of you. One is the profile you edit and is displayed to other users. The other is the profile that the platform creates by including personal attributes and traits that you did not explicitly provide, but were inferred from your digital traces and activities on the platform.  % clarify

Table~\ref{tab:models} summarizes the characteristics of the models we discuss in this section. The table also includes digital voodoo dolls which we discuss in Section~\ref{sec:voodoo}.
\subsection{Identifying models} 
 We start with what we call  {\em identifying models}. Biometrics are used to model the physiological characteristics of a person.  States use {\em personal data}, including but not limited to biometrics, to build a model of their citizens. What constitutes personal data is typically regulated by law. A personal identity document, or a personal record, can be seen as a model of a person. We use personal records to keep track of the students enrolled in a university, patients that have received treatment etc. 

What all these identifying models  have in common is that the model is intended to uniquely identify the person with which it is associated. In other words, there is one model to one institution or a one-to-one relationship between the set of models and the set of persons. Strong time-invariant uniqueness is guaranteed for identity models. Furthermore, they are not transferable. For example, a passport belongs to exactly one person until the end of time and this is guaranteed by including the right type of data in the model. The data in identity models cannot be changed without the joint consent of the user and the institution that creates the model. 

Clearly, the purpose of identifying models is to identify an individual within the institution allowing adequate rights and obligations to be associated with that person.

\subsection{User accounts and verified user accounts}
A different type of person models are {\em accounts}, also called user accounts. An account is a model that is created by a person, within the context of some service. We thus have email accounts, social media accounts, phone service accounts, etc. This type of model is weakly permanent and weakly unique to a person. Namely, although the account can only be associated with one person at a time, it can be transferred to someone else. In contrast, identifying models, such as biometric data or social security numbers for example, cannot be transferred. 

User accounts are models, are one-to-many relationship with a person, in the sense that one person can have more than one account at the same time, but one account is not (intended to be) shared by more people.  The uniqueness of these models is time-variant. For example, two email accounts or phone numbers can belong to the same person, but after the person has closed the account, that same account or number can be allocated to someone else.  Accounts are transferable - they can be but are not necessarily associated with an identity model.  

Let us call {\em verified} the accounts that are necessarily associated with an identity model. The term ``account verification" was was introduced by Twitter in June 2009 \citep{Kanalley2013}. However, the process of verifying the identity associated with a particular account predates social media - a person might sign up for a service and open an account but that relationship is only valid once the institution has verified the identity and associated properties of that user.  
 
Both user accounts and verified accounts can be created by the user, and the user is the one that has edit control over their account. The account itself can be terminated unilaterally by either party.

The purpose of (verified) user accounts is interaction with the institution and interaction among different users in the context of that institution. For example, you can only send an email to a person that has an email account. 

\subsection{User models and user profiles}
 Human-computer interaction (HCI) is a field that studies the interaction between a person and a computer in sociotechnical environments. HCI was necessitated by the increase of complexity of computer systems and the complexity of their uses and purposes. Specifically, HCI is concerned with enabling computer systems to provide its intended users with the interaction experiences fitted for that user's knowledge and objectives, since the users with different knowledge and abilities can interact with same system pursuing different goals.

A {\em user model} is a data structure that represents a user in a  computational environment, typically by representing the relevant (for the environment) characteristic of that user. For example, characteristics such as age group, language, most used application, dominant hand all can be part of a user model. In other words, a user model describes what is of interest about a user to a specific computational environment. A {\em user profile} is the representation of a specific user. In our example, this would be for instance 25-35, English, Overleaf, left-handed. 

Neither user models nor user profiles are unique: many persons are associated with the same user model and  may be associated with the same user profile. The features that comprise a user model are selected by the designers of the computational environment. The information that is used to ``fill'' the user profile is elicited from the user either directly or by means of analyzing their digital footprints. For example, web-pages are displayed differently for different user models accessing them based on for example, location, bandwidth and type of device accessing them.

 User modeling was developed when interaction with computational systems was unidirectional: the computational system offers information and the user consumes it. In this sense, a user model is a model associated with a particular computational system. In contrast, web 2.0. systems are bidirectional - the user now not only consumes information, but they also provide information to the system.
 
Web 2.0 is a characterization of, initially web pages, but later also internet based services, and computational systems in general, used to identify, among else, that the system admits user-generated content and supports interoperability \citep{BlankR:2012}. Web 2.0 changed how user models are constructed by making it easier for system designers to infer the characteristics and intentions of their users. Specifically, the interoperability aspect of Web 2.0 systems allowed for user features and personal information to be obtained using digital footprints. 
Consequently, unlike identity models and account models, user models can be created without the explicit consent of the person which they model. 
The practice of allowing  a person to modify or choose their user model is called {\em open user modelling}. 

The purpose of user modeling is {\em personalization}. User modeling has been developed to provide systems with the ability to ``say the `right' thing at the ``right" time in the ``right" way" \citep{Fischer2001}.  
 
\subsection{Look-a-like models}
In a pre-Web 2.0 socio-technical system, the user chose the system they would interact with. A user model and a user profile are constructed to adapt the computational system operation to the user that has engaged with it. The bidirectional nature of Web 2.0 and the rise of social media as an attention capture platform \citep{Wu2016}, has allowed for a bidirectional engagement -  the users can be ``targeted'' for a particular service or product. While personalization is the process of choosing the right service for a person, targeting is the process of choosing the right person for a service. 

 Lookalike Audience is a service initially offered by Facebook in 2013, in which advertisers choose aspects of a user model, such as age, gender, location, language etc and pay for their content to be shown to persons which can be associated with the chosen look-a-like model. Facebook then sought users that fit a model, rather than seeking a model to represent their users. An advertisement is shown to all people who match a particular set of properties (e.g., female, 15-20 years, has visited dating sites, has searched for vegan recipes). 
 
 A related practice that predates lookalike audiences is collaborative filtering. Collaborative filtering is a method used in recommender systems \citep{Ricci2011}. This is the process of  identifying items to be recommended to a user either by determining if people similar to that user liked the item in question, or if the user liked, in the past, items similar to the item in question. In either case, the related preferences are obtained by aggregating the information from many users or items (hence collaborative). The underlying assumption is that if two people share the opinion regarding an item or a service they have both experienced, then they are more likely to share the opinion regarding a related item. In effect, a user is matched to a person model. An item is recommended to all people who match a particular set of properties - those of people who have in the past liked the item. 
 
 A look-a-like model is created for targeted content exposure. Targeted content, as a purpose for creating look-a-like models itself can have several purposes. Apart from getting the ``best attention'' for a product or service, it can also be used to influence particular people by exposing them to carefully curated content. 

 A look-a-like model is built based on an identified pattern of characteristics among the users of a sociotechnical system that is linked to a behaviour of interest. One user can ``fit" with several look-a-like models.  Look-a-like models are built to be applied to many people. A lookalike model is under the control of the enterprise that commissioned it. Clearly, a look-a-like model can be created without the knowledge and permissions of the users. Furthermore a person may not necessarily be informed that they are matched to a particular look-a-like model - they may be aware only of the consequences of this match. 
 
 A person can sometimes influence the look-a-like model to which they are matched by changing their user model in the same system. For example, if one does not want to be targeted by advertisements for women they can miss-represent their gender. However, given the extent of digital footprints and the inferences that can be made about a person, a look-a-like model does not need to pool from user models to identify the users that match it. For example, if your friends use gender definite pronouns when they address you in their posts, your gender can be assigned to you from these posts rather than from your user profile.  This approach can cut off the control a user has with respect to being targeted. 

%cybernetic categorization from piece Marija sent me!! Or even a cybernetic ontology. From teleological to statistical/computational ontology 
%"We are effectively losing control in defining who we are online, or more specifically we are losing ownership over the meaning of the categories that constitute our identities. - Game of Throne correlation example. Can't control my context.
 
\section{Digital Voodoo Dolls}\label{sec:voodoo} %JON TO DO 1) all I do is interact 2) I don't need to interact with it directly, if I interact with people who interact with it. User models of Facebook pixel or like button %add in sign-posting. . Prediction model: "the user has the intention to do this". Use reductio ad absurbum for voodoo model (Black people and prison calling cards). Distinction between treating all users as the same until they USE the system, versus making an initial prediction. 
 
In this section we define the a digital voodoo doll as a model of a person and contrast its properties with the properties of the models we already introduced. The choice of ``voodoo doll" as term around which we build our concept is fairly intuitive. Nevertheless we briefly discuss the idea of voodoo dolls and justify our usage of the term. Tristan Harris has mentioned voodoo dolls in the context of social media penalisation:  % in Section ``Voodoo dolls"%~\ref{sec:native}.

 \begin{quote}
   [I]n the moment you hit play, it wakes up an avatar, a voodoo doll-like version of you inside of a Google server. And that avatar, based on all the clicks and likes and everything you ever made ``those are like your hair clippings and toenail clippings and nail filings that make the avatar look and act more and more like you so that inside of a Google server they can simulate more and more possibilities about if I prick you with this video, if I prick you with this video, how long would you stay?" And the business model is simply what maximized watch time.\footnote{\url{shorturl.at/akpq1}}
 \end{quote}

Voodoo dolls are magical objects which voodoo practitioners create to influence others; they function by virtue of an entanglement between that person and the aspects of their likeness captured by the doll. Anthropologists define two general kinds of voodoo dolls: {\em homeopathic} and {\em sympathetic}. Homeopathic voodoo dolls are created from materials actually gathered from an individual in question. The doll stands in for the person whom one wishes to manipulate by a part of them which is attached to the doll: a clip of hair, a scrap of clothing, etc. Sympathetic dolls are created by molding a doll into the image of the individual one wishes to influence. The doll magically stands in for the targeted person by virtue of the fidelity of the modeled image.

Digital voodoo dolls are models of persons which web platforms manipulate in the hopes of achieving an underlying change in the behavior of the person being modeled. In contrast to magical voodoo dolls, these digital voodoo dolls are not necessarily the intentional product of a sole creator, but are an emergent product of a socio-technical system. Digital Voodoo Dolls are distinct from various kinds of user models in that the data they aggregate, and the choice architectures they influence, may dwell in data flows beyond the control of any single internet actor.

\begin{figure}
    \centering
    \includegraphics[width=0.7\textwidth]{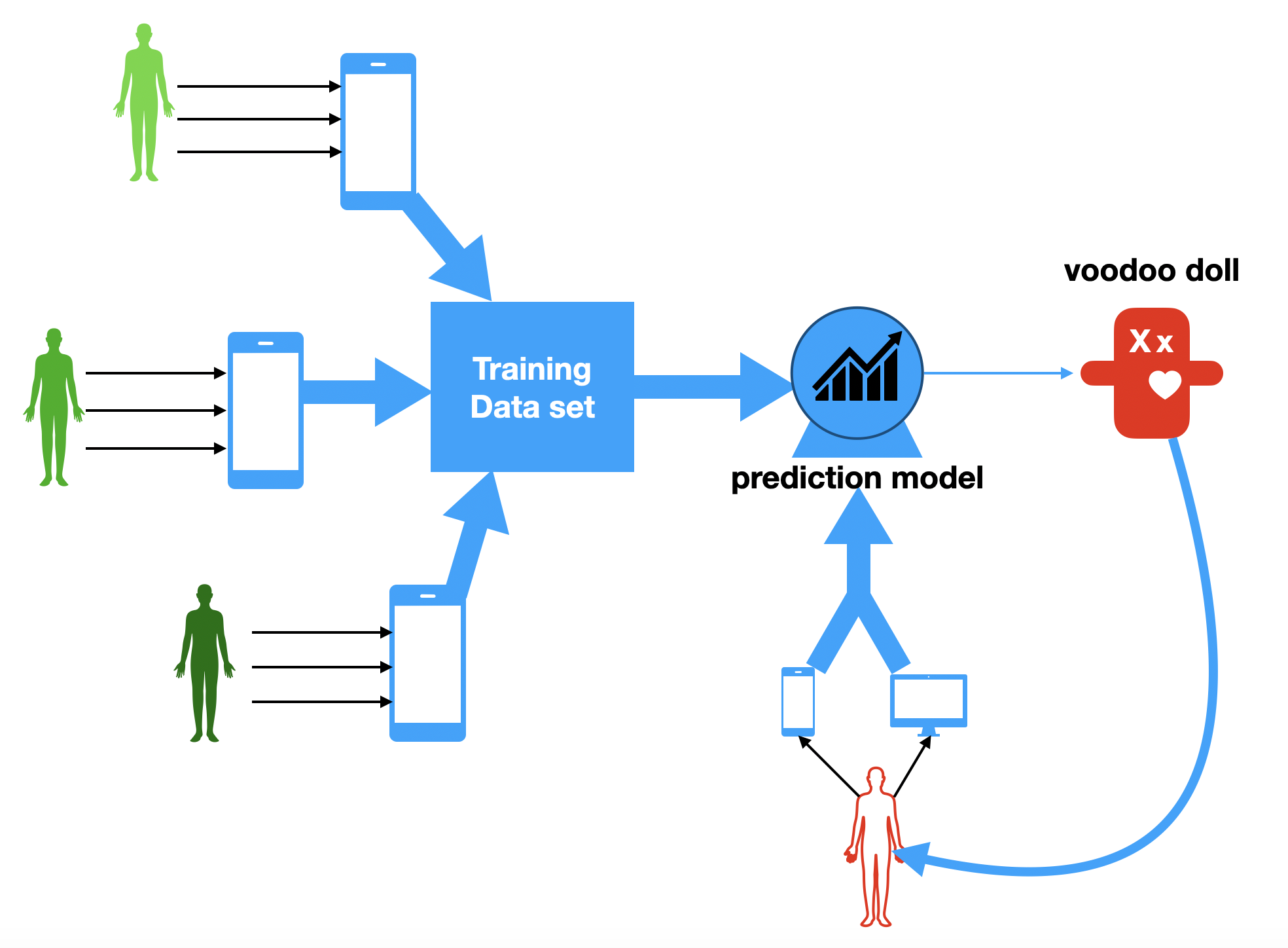}
    \caption{The creation of a digital voodoo doll}
    \label{fig:doll}
\end{figure}

Let us begin by describing how a digital voodoo doll is created. Figure~\ref{fig:doll} shows a diagram of the process. Features of many people are used to construct a training data set that is used by one or more prediction models to predict a future behaviour or choice of a person. The prediction model is used on features of a specific person of interest to predict that person's behavior, choice or interests. When a new user accesses the system, the prediction model will use all available information (including, or perhaps especially, technical data on geography and device-type), matched against the preferences of those with similar signatures whose data it already has, to construct an initial predictive model of that user's preferences. The resulting prediction, together with the person's features, create the digital voodoo doll construct. % emphasize it will pull this data from wherever.

What distinguishes modern digital voodoo dolls from look-alike models is that they pull data programmatically from multiple socio-technical systems, as opposed to relying solely on past data from any particular platform. Google and Facebook pull data from a staggering network of tracking systems embedded in pages that sell Google ads or use the "Facebook Pixel" service; this data is then conveyed to ad brokers both on and off of those systems. Disparate third-party vendors collect or buy user data, pool it into identifying profiles, and re-sell it to ad sellers, marketing firms, political survey companies, etc. (and, of course, when it is used to sell ads via the Facebook or Google networks, this information is also added to those systems). Companies producing social scoring systems, especially tied to risky "offline" behaviors like lending or home rental, purchase this data, match it to other public and private data, and resell it as risk scores or social scores of different kinds. All of these activities are embedded as services at every part of the contemporary digital revenue chain. Both individual data and data on look-alike users are collated to produce recommendations or scores that underlie modern choice systems. It is increasingly difficult for an individual user to understand where and how any system she interacts with has accessed their data, or whether recommendations are being made based on discrete individual data or predictions based on other, similar users. This had led to a feeling of uneasiness about modern recommendation systems \citep{stern_facebook_2018}.

\subsection{Person-Institution Relationship}

A voodoo doll is unique to a person because it is generated by feeding the features of a specific person to one or several prediction models, including identifying information (such as IMEI\footnote{International Mobile Equipment Identity (IMEI) is a unique 15-digit code that precisely identifies the device with the SIM card input}, IP, device, browser, and geographic signatures). Digital fingerprinting and the use of sticky tracking protocols underlie the ideal goal of providing a one-to-one person-to-voodoo-doll relationship, even if the reality may ensure some bleed-over of information from other people becoming associated with your account; this may be trivial in the case of a family member using your Netflix account, but consequential in the case of, for example, the data of someone with the same name as you becoming added to a social scoring system.

\subsection{Permanence}

A voodoo doll is dynamically generated by a recommendation or scoring engine; however, the ability to generate the voodoo doll is permanent as long as a prediction model can be built and features of the person being modeled are available. Assuming the same data flows or third-party data are available, the recommendation system can readily re-build my voodoo doll. For instance, if I instruct Facebook to delete all of my information and delete my account, it will do so. However, as soon as I start browsing the internet again, millions of websites using Facebook Pixel will begin re-constituting a profile on me, so that if I once again create a Facebook account, a thorough model of me will be instantly available. 

\subsection{Can create/delete} 

 What all the discussed models in the previous section have in common is that they are ``institution specific''. Namely, they are created within the scope of one institution, or one sociotechnical system, using the information provided within that system to serve the purposes of that system. Different systems would create different user models for their purposes and the sharing of the models is formally specified by agreements between the institutions and the person who is modeled. A digital voodoo doll however exists as a cross-institutional construct: the information from several sociotechnical systems can be combined to construct it without those institutions having a formal agreement to share information. For this reason, a voodoo doll model cannot really be deleted because the data used to generate the voodoo doll model continues to exist and can be reused to re-create the model. 

 A particular instantiation of a voodoo doll can only be deleted or edited by the institution that has created it. This can be done without making aware the person being modeled and that person has no way to influence at all which voodoo doll they will end up with since the data used in the voodoo doll's creation is past digital traces.

\subsection{Can edit}

Because the voodoo doll is ultimately a cross-institutional construct made of data flows and social scores that are aggregated programmatically to build option architectures, the underlying construct is not owned by, and cannot be edited by, any one institution or person in particular.  %tk 

\subsection{Purpose}

The voodoo doll model can be used as a representation of a person that is analyzed to assess to which extent can the associated person be trusted, to assign other personality traits or profiles to this person, to infer possible future behavior in given contexts, or to assign a ``social score", ``risk score", or a ``client score" to them. Voodoo dolls combine the one-to-one goal of an identity model with the "thick" social and personality description of data-driven modeling. This combination renders individuals legible and predictable to digital systems, in ways that permit improved ad sales, improved performance of recommendation systems, price discrimination, service discrimination, and other advantages for digital companies. The advantage for companies is that deriving identifying information rather than asking users for it improves system performance even in cases where users cannot or will not sign up for accounts and provide information because it would disadvantage them in some way. In addition, user data that is derived from fingerprinting and not technically identified with personal information is more liquid and comes with fewer fiduciary responsibilities in a post-GDPR world. Thus, companies can buy and sell "anonymized" location data far more readily than not-anonymized data, and companies that purchase and use it face fewer legal responsibilities, even if the data can be trivially associated with an individual through voodoo doll modeling. Institutions get the benefits of identifying models without any of the costs.

Voodoo dolls are emergent because the data used to create these models is harvested and in no way directly elicited or pursued. They are persistent because the data used to build them exists by virtue of a person's existence in the digital world. The only way a person could change a voodoo doll model is by changing their behavior and personality in the past. Yet, a person is directly affected by the voodoo doll model associated with them, (e.g., services can be made unavailable). This complete severance of access to the model by the person modeled can be a source of discrimination and violation of other personal rights and freedoms.

\section{Ethical concerns} \label{sec:concern} %Cash out Theory section. What does it mean? What should we do?
AI ethics is a new interdisciplinary research area motivated by, among other factors, the increased involvement of artificial intelligence (AI) in the evaluation of people's abilities. The use of digital voodoo dolls raises numerous ethical issues in AI ethics because this use directly impacts a person's right to privacy, autonomy and freedom. We discuss them in this section.

Privacy is studied in the context of security \citep{Dwork2006} where the problem is the endangering of the safety of an individual by exposing secret data such as their social security number. Within AI ethics, we are typically concerned with a different interpretation of the concept of privacy, as in the phrase ``invasion of privacy",  where the concern is that an intrusion into the personal life of a person is made. 

The digital voodoo dolls share the concern for violations of privacy with the general possibility to collect data in the form of online digital footprints and as behavioral records from consumer electronics \citep{PrivAI}. The privacy concerns have been particularly exasperated by the fact that these new possibilities have also led to a data-driven economy, often termed surveillance capitalism \citep{Zuboff2019}, that is based on the digital agglomeration of information about individuals. Surveillance capitalism has added motive to the existing possibility to infringe upon the privacy of people. 

In AI ethics privacy often defaults to differential privacy, but the two concepts are not the same \cite[Chapter 1]{KearnsR2019}. Differential privacy studies how data sets can be used without revealing information about individuals described in them. However, in digital voodoo dolls the privacy concern is for the person who is modeled by the doll, not only those whose data have been used in the doll's generation. Differential privacy offers no protection to the modeled person. In the words of \citet{KearnsR2019}: ``Remember that, by design, differential privacy protects you from the consequences of what an observer might be able to learn specifically because of your data, but does not protect you from what an attacker might be able to infer about you using general knowledge of the world''. That ``general knowledge of the world" can be data about you compiled through third-party data brokers and then fed to a prediction model. Even if you had some control over whether data will be collected about you, you do not have control over who that data will be sold to and how it will be used from that point on.

Individual autonomy is typically defined as the ability to construct one's own choices, and to have the freedom to make one's own decisions and perform actions based on these decisions \citep{Brey2006}. The possible infringements on the autonomy and freedom of a person are far more specific in the digital voodoo doll practice. Some of these have been expressed in the numerous popular science articles
\footnote{\url{https://www.nytimes.com/2020/05/28/business/renters-background-checks.html?smid=nytcore-ios-share}}
\footnote{\url{
https://www.engadget.com/2020-01-17-your-online-activity-effectively-social-credit-score-airbnb.html}}
\footnote{\url{
https://algorithmwatch.org/en/schufa-a-black-box-openschufa-results-published/
}}.

In all person models that precede the digital voodoo dolls, a person has a certain amount of autonomy to choose not to use the service that makes the model. Digital voodoo dolls take away this freedom: anyone who has your digital ``materials" can make a digital voodoo doll and use it without your approval or awareness to change the reality you experience. 

Digital voodoo dolls are built so that services offered and provided can be tailored to that model. This means that when there is an error in the model, the person modeled will be unjustly affected. All person models can contain errors. What is particular for models that are not accountable to the user is that they strip that user from the ability to identify the error and take actions to have it corrected.  If a person does not know that the digital voodoo doll exists, they will not be able to understand why they are experiencing a particular range of options or types of services and with that, how to improve one's own circumstances.  

In contrast, the awareness that a digital voodoo doll is used imposes a new type of limitation to freedoms and autonomy. A person can be told that it is their undesirable past conduct that has led to diminished options and services. This can serve to nudge or sludge them \citep{Thaler2018} towards a particular desirable (for the doll creator) behaviour and choices. At which point does a nudge (or a sludge) become an infringement of autonomy? 

Concerns regarding violations of freedom and autonomy have been raised in the context of ambient intelligence. Ambient intelligence is an environment in which devices perform tasks by responding to the presence of people using intelligent user interfaces to profile their users, that is to personalize and automatically adapt to a user's behavior patterns, and possibly anticipate their intention and requirements. By design, ambient intelligent devices are ``invisible'' and users may not be aware that they are interacting with an intelligent device \citep{Tavani2012}. This invisibility by design is what ambient intelligence shares with digital voodoo dolls. 

Ambient intelligence and in general in society has been continuously escalated in the past fifteen years \citep{Brey2006,Moor:2006,Noble2018,Oneil2016,NguyenYC14,Brey2006}. The concerns revolve around issues of physical safety, privacy, invasion of privacy, limitation of autonomy and freedom.  \citet{Brey2006} discusses the negative impact on autonomy and freedom, in addition to the impact on privacy. He outlines three ways in which a loss of control can occur in an ambient intelligence environment: actions can be performed that do not correspond to the needs or intentions of the user, an increased effort can be required of the user to bypass, switch-off or disengage the artificial agent's preferred course of action, and when the operation of an intelligent agent is not only governed by the interest of the user but possibly conflicting interests of third party stakeholders are also engaged.  All of these three ways of control loss are applicable to digital voodoo dolls as well. 
However, in an ambient intelligent setting, the user still has the choice to physically interact with that environment and enforce their will - leave the building. The digital voodoo doll cannot be ``left".

Lastly, the creation of digital voodoo dolls depends on the availability of large amounts of data used to construct prediction models with machine learning. As with the machine learning prediction about people's features in general, here too we encounter issues of fairness and bias - if there has been a past correlation between a person's features and their behaviour, these will be embedded in the prediction model.

There are currently more than twenty different fairness measures \citep{Mehrabi2019}. Statistical notions of fairness, also called group fairness, take the approach of identifying a small number of protected demographic groups and requiring that (approximate) parity of some statistical measure across all of these groups is maintained. Individual notions of fairness are focused on individuals rather than groups requiring that ``similar individuals should be treated similarly''.  Similarity is a task-specific metric determined on a case by case basis \citep{Dwork2011}. 

Bias mitigation efforts are currently focused on improving the measures of fairness \citep{Mehrabi2019}. These measures do not capture the unfairness that can be built in a digital voodoo doll because the doll is a personal model.   \citet{ChouldechovaR20} argue that ``statistical definitions of fairness do not on their own give meaningful guarantees to individuals or structured subgroups of the protected demographic groups. Instead
they give guarantees to `average' members of the protected groups."  

 %\todo[inline]

\section{Possible paths to responsible voodoo}
In this section, we discuss two approaches towards empowering users to regain control over their digital voodoo dolls. Both approaches seek to attain the same goal: establish algorithmic accountability for digital voodoo dolls. Algorithmic accountability is a relationship between an actor and a forum in which the actor has to justify or explain some aspect of the properties of an algorithm to a forum that has the power to pose algorithmic requirements and sanction the actor if the algorithm does not satisfy these requirements \citep{Wieringa2020}. Intuitively, algorithmic accountability is about ensuring a person's ability to change the behaviour of an algorithm with which they are interacting. A core problem with the creation of digital voodoo dolls is that algorithmic accountability is difficult to establish. This is due to the persistent nature of the digital voodoo doll - its existence is proof that it can be recreated. 

The first approach is based on establishing indirect user control: not over the digital voodoo doll itself, but over the data that enables it.  By design, the person affected by the use of a doll has no power over the actors who created the doll to persuade or force them to not use it. The only option is to explore how a person can achieve that a digital voodoo doll is not created in the first place. Some regulatory progress has been made despite the complexity and opaqueness of online tracking mechanisms which make it impossible to get a sense of the data that was automatically collected, who has access to it, and most importantly what purposes it serves. Digital voodoo dolls are the embodiment of this complexity and opaqueness and the user that wishes to curb their use can benefit from these regulations. 

The second approach is the approach of netizen disobedience.  Regulation can be used to control how much data as possible to collect and store about me (a user). However, there is a rise in technical solutions that work to make my data traces unusable.

\subsection{Regulatory approach}
Here we explore to which extent can existing regulation be used to establish accountability. Digital voodoo dolls exist by virtue of digital data. If a person has control over the existence of digital data traces they leave behind, they indirectly can control the fidelity of a possible digital voodoo doll. 

Legal scholars have recognised that the permanent and timely unlimited storage and distribution of a person's digital traces is most likely not purely beneficial to them. This has been the basis for establishing ``the right to be forgotten". The right to be forgotten is the  right of a person to require that information about them be removed from internet searches and online accessible directories \cite{Rosen2012}. This right has been included as part of the General Data Protection Regulation (GDPR) of the European Union (EU) as Article 17 ``Right to erasure"\footnote{\url{https://gdpr-info.eu/art-17-gdpr/}}. 

The GDPR includes Articles 16, 17, 21 and 22 which address, respectively, the right to ask for incorrect information to be corrected, to have personal information erased, to object to processing of personal data, and the right not to be subject to a decision based solely on automated processing, including profiling. 
Moreover, as we were preparing this submission, the EU has revealed a proposal\footnote{\url{https://digital-strategy.ec.europa.eu/en/library/proposal-regulation-laying-down-harmonised-rules-artificial-intelligence-artificial-intelligence}}  for a new policy on the regulation of AI systems that also accounts for social scoring and automated personality inferences. Once ratified by the European parliament, this regulation could be the first of its kind, regulating the creation and existence of some concepts outlined in this paper.

The California Consumer Privacy Act (CCPA) largely built on the framework of the GDPR. Both acts broadly aim to protect the consumer, but the regulatory approaches do slightly differ. Compared to the CCPA, the GDPR's coverage is broader. The CCPA includes the pillars of the GDPR (Articles 16, 17, 21 and 22) with one notable expectation: the right to request incorrect information to be changed upon request. The scope of the CCPA is also significantly more limited as it only applies to larger businesses that reach a certain gross revenue threshold (25 million United States dollars annually), generate half or more of their annual revenue from selling personal information, or handle the personal data of more than 50,000 California residents. Whereas the GDPR applies more broadly to companies that either operate within the European Union or handle the data of European Union citizens. 

Compared to the CCPA, Brazil's General Data Protection Law (Lei Geral de Prote\c{c}{\~a}o de Dados Pessoais or the LGPD) is notably more similar to the GDPR. The LGPD contains the rights of the Articles 16, 17, 21 and 22 of the GDPR and is broadly applicable for companies operating within the territorial jurisdiction of Brazil and offers an additional protection for all Brazilian nationals \cite{Koch2018}. 

Citizens and users can only fully benefit from legal protection if they are aware they have the right to it. We have had many years of practice in a system of human service providers and users, and comparatively a very short time of transition to a system in which digitization has joined this relationship. Digital voodoo doll creators are enabled by people not aware of the legal protection they are entitled to. There is, of course, a limitation to how much the right to be forgotten type of regulation can accomplish.

The regulation does not necessarily include the right to have data removed from data collections. We need further legal frameworks to enforce that in consent forms the right to use the data to provide a service and to sell/copy the data or use it to create latent representations of people is distinguished \citep{Wijesekera2017}.  

Another requirement to request the deletion of data on a specific person is the knowledge that this data exists in the first place or that you are contributing to it being collected \citep[e.g., through privacy nudges, ][]{Almuhimedi2015}. Currently, large quantities of digital information on individuals are hidden away in private online repositories that are only made available to affluent buyers. Thus, although people could request their personal data from companies, the uncontrolled and progressing distribution of digital footprints across the web renders this endeavour futile if new copies can be created faster than they can be deleted. Hence, to effectively achieve the deletion of the voodoo doll, internationally effective legislation and mechanisms need to be developed to both allow to find the voodoo doll and to delete all its copies and fragments.

Once legal frameworks are in place, the technical act to remove personal data from search results and the web in general could be achieved in a similar form as unique identifiers are currently used to combine these data to create most personalized ads.

While there is clearly progress in the direction of regulatory protection, having regulation in itself does not address the issues of digital voodoo doll creation and use. On one hand, the number of digital data collection and processing violations clearly overwhelms any singular country's capacity to investigate and prosecute these violations. On the other, anyone in the world can be a subject to digital voodoo doll modeling, while regulatory protections are offered only in some jurisdictions. We next discuss the means available to all netizens regardless of where geo-politically they are. 

%to have private information about a person be removed from Internet searches and other directories under some circumstances. The concept has been discussed and put into practice in both the European Union (EU) and in Argentina since 2006.[1][2] 
%This is why modern regulatory action incorporates peoples' rights to demand the deletion of their online history from companies (e.g., GDPR, CCPA, GDPL)
\subsection{Netizen disobedience}

Some of the data traces we leave behind are choices we make while online: engage with a product recommendation or not, spend different amount of time composing an email to a person we care deeply about compared to a person we do not, stop scrolling when we see an amusing advertisement. It is this choice from essentially a set of options, that distinguishes us. If each of us chooses no option or if each of us chooses all options, then we cannot be discerned in the crowd. There are numerous technological solutions developed that enable us to do just that. 

Ad-blockers are a simple example of such a mechanism. They are small browser plugins that detect and omit advertisements when the browser loads a website. Ad-blockers eliminate the exposition of the user to advertisements on websites and therefore eliminate the possibility to react differential to those. In a similar way, browser-plugins can be used to block tracking code in websites.

A slightly different effect could potentially be achieved through obfuscation. Obfuscation does not prevent people from being tracked but rather changes the data that is produced in the process. In that sense, people's input data could be enriched or blurred with additional data to mask a person's true behavioral traces with fake ones. Text input could for example be altered in an intermediate layer and be enriched with different words and embellishments or reproduced in a completely different form (e.g., by using generative language models).

Similar to the saying that clothes make the man, your digital "appearance" makes you.
Hence, advancing the idea of data obfuscation, it would even be possible to artificially enhance people's digital footprints in the desired way by adding or omitting selective types of information. For example, specific machine learning models \citep[e.g., advancing the idea of counterfactual explanations,][]{Dandl2020} could be created that would predict the smallest change necessary in your digital footprints for you to appear in a different light (e.g., buy a book, visit church). This could in turn could systematically change your voodoo doll and have significant implications on decisions that are being made based on it (e.g., credit scoring). Finally, these obfuscation methods could be used in combination with multiple aliases to distribute enhanced digital footprints across different artificial persona. While these ideas dance on the border of legality and would require sophisticated reverse engineering of AI decision systems, it could create a new market of online masking or digital self-optimization. Moreover, these approaches seem increasingly warranted from an ethical point of view, in particular in societies where your digital voodoo doll could have severe implications on your daily life \citep{Wang2018, Andersen2020}.

There inherently exists an asymmetry of power between the people who build AI systems and the people who interact with them. A feature of AI is that it offers a personally tailored approach - adapts to the person it interacts with. But this directly leads to a user not having access to information on how an algorithm works for someone else. Without this information the user can suspect, but never establish the bias of the algorithm. How many users does it take to acknowledge that an AI system is for example unfair, or that a digital voodoo doll has been used to tailor services? 

The Internet makes the digitization of our lives possible, but it also helps us broadcast our insights and connect with people who can validate our experiences. 
Taking advantage of this is to create a public forum in which experiences can be shared. Numerous such forums exist to record fraudulent internet-based services and practices.  There is now an initiative to record AI incidents, \url{https://incidentdatabase.ai/}. The downside of these forums is that there is no framework to establish who and when should act to investigate, confirm and mitigate misconduct.   

\section{Summary}
All models are wrong, but some are useful. We have argued that there is a new person model that is being used, the digital voodoo doll, which when wrong is not only not useful but also posses serious ethical concerns. We have proposed a conceptualization of the digital voodoo doll. By comparing it to existing person models that are used in our socio-technical societies, we demonstrate that indeed this data structure is new and its features are unique. We draw attention to the concerns that digital voodoo dolls introduce with the aim of ensuring that different stakeholders in society are prepared for and work towards facing those concerns. 

Here we offer an analytical analysis of the digital voodoo doll concept. In future work, we intend to prepare an empirical analysis. This includes exploring the prevalence of digital voodoo doll use, by exploring the motive and opportunity. There opportunity here is the exploration of who can collect personal information on demand, while motive is the exploration of the incentives an agent would have to create and use digital voodoo dolls. 

Digital voodoo dolls are characterized by being completely out of the control for the person they represent. By design, direct control is not possible. We briefly discussed possible approaches to empower a person to indirectly assume control by preventing a digital voodoo doll from being created. Numerous avenues for future work are open following both our approaches: establishing legal frameworks to guarantee the rights of people and building technological tools that empower people to opt out of being traced. 
Lastly, and most importantly we want to stress that this work should not be seen as a general condemnation of modern, data-driven processes in our societies. In particular, this is not our intention, because the increasing possibilities to digitally represent and model natural phenomena have led to radical improvements in the understanding of our world. Moreover, data tracing and artificial intelligence can help us to tackle the most challenging problems of our times (e.g., COVID-19 tracking). Our intention is rather to emphasize the importance to monitor and illuminate developments in this area from different points of view to ensure that everyone can benefit from AI fairly \citep{Taddeo2018}. 

\bibliographystyle{ACM-Reference-Format}
\bibliography{Bibliography}
\end{document}